\documentclass[a4paper,twocolumn,pra,%
amsmath,amssymb,showpacs,floatfix]{revtex4}
\usepackage{graphicx,epstopdf}

\newcommand{\ket}[1]{\left|#1\right\rangle}
\newcommand{\tr}{\mathrm{tr}}
\newcommand{\A}{\mathcal{A}}

\graphicspath{{images/}}

\begin{document}

\title{Optimizing for an arbitrary perfect entangler: I.\ Functionals}

\author{Paul Watts and Ji\v{r}\'i Vala} 
\affiliation{Department of
  Mathematical Physics, National University of Ireland, Maynooth, Co.\ Kildare,
  Ireland} 
\affiliation{School of Theoretical Physics, Dublin
  Institute for Advanced Studies, 10 Burlington Road, Dublin, Ireland}

\author{Matthias M. M\"uller}
\affiliation{Center for Integrated Quantum Science and Technology,
  Institute for Complex Quantum Systems, Universit\"at Ulm, D-89069 Ulm, Germany}

\author{Tommaso Calarco}
\affiliation{Center for Integrated Quantum Science and Technology,
  Institute for Complex Quantum Systems, Universit\"at Ulm, D-89069 Ulm, Germany}

\author{K.\ Birgitta Whaley}
\affiliation{Department of Chemistry,
University of California, Berkeley, California 94720, USA}

\author{Daniel M.\ Reich}
\affiliation{Theoretische Physik, Universit\"{a}t Kassel,
  Heinrich-Plett-Str.\ 40, D-34132 Kassel, Germany}

\author{Michael H.\ Goerz}
\affiliation{Theoretische Physik, Universit\"{a}t Kassel,
  Heinrich-Plett-Str.\ 40, D-34132 Kassel, Germany}

\author{Christiane P.\ Koch}
\affiliation{Theoretische Physik, Universit\"{a}t Kassel,
  Heinrich-Plett-Str.\ 40, D-34132 Kassel, Germany}

\date{\today}

\begin{abstract}
Optimal control theory is a powerful tool for improving figures of
merit in quantum information tasks.  Finding the solution to any
optimal control problem via numerical optimization depends crucially
on the choice of the optimization functional.  Here, we derive a
functional that targets the full set of two-qubit perfect entanglers,
gates capable of creating a maximally-entangled state out of some
initial product state.  The functional depends on easily-computable
local invariants and uniquely determines when a gate evolves into a
perfect entangler.  Optimization with our functional is most useful if
the two-qubit dynamics allows for the implementation of more than one
perfect entangler.  We discuss the reachable set of perfect entanglers
for a generic Hamiltonian that corresponds to several quantum
information platforms of current interest.
\end{abstract}

\maketitle

\section{Introduction}
\label{sec:intro}

Entanglement between quantum bits plays a fundamental role in quantum
information processing.  It can be generated between two qubits by
suitable operations from the Lie group $SU(4)$.  The geometric theory
of $SU(4)$ formulated by Zhang et al.~\cite{ZhangPRA03} provides a very
useful classification of two-qubit operations in terms of their local
equivalence classes.  These are uniquely characterized by three real
numbers known as local invariants~\cite{Makhlin}.  Each local
equivalence class contains all the two-qubit gates which are
equivalent up to single-qubit transformations and is characterized by
a unique nonlocal content and thus has unique entangling capabilities.

The geometric theory has recently been combined with optimal control
theory by M\"uller et al.~\cite{MuellerPRA11}.  Specifically, using
the local invariants which uniquely characterize local equivalence
classes, the optimization target was expanded from a specific unitary
operation to the corresponding local equivalence class.  This
considerably relaxes the control constraints.  The ensuing
optimization algorithm~\cite{MuellerPRA11,ReichJCP12} allows for
identifying those two-qubit gates out of a local equivalence class
that can be implemented, for a given system Hamiltonian.  The
algorithm can be employed to determine the quantum speed
limit~\cite{TommasoPRL09}, i.e., the fundamental limits for a given
two-qubit system in terms of maximal fidelity and minimal gate time.

Here, we further explore the potential for quantum information
processing offered by the combination of geometric theory and optimal
control.  Our starting point is the characterization of  perfect
entanglers provided by the geometric theory.  Perfect entanglers (PEs)
are nonlocal two-qubit operations that are capable of creating a
maximally-entangled state out of some initial product state.  In
particular, we define a function to uniquely and easily identify
whether
a two-qubit operation is a PE.  Since this function is given in terms
of the local invariants, it can be easily incorporated into the
optimal control functional used in Refs.~\cite{MuellerPRA11,ReichJCP12}.
This allows us to expand the optimization target to the full set of
PEs, which corresponds to half of all local equivalence classes.
The optimization functional may be thought of as
measuring the ``minimal distance'' between the gate $U$ and the subset of
matrices in $SU(4)$ which are PEs.  It is zero for a PE and positive
otherwise.  The
functional is remarkably easy to compute for any matrix, requiring
only elementary algebra.

Optimization targeting the set of PEs will proceed along a path in the
Weyl chamber, i.e., the reduced two-qubit parameter space, if the
system dynamics allows for 
implementation of only a single local
equivalence class containing a PE.  However, our approach is most
useful if more than one local equivalence class containing a PE can be
reached.  Optimization will then explore a larger portion of the Weyl
chamber.  We therefore also present an analysis of the reachable set
of local equivalence classes, considering a generic two-qubit
Hamiltonian that models superconducting qubits.  The application
of our optimization approach to
examples is presented in the sequel to this paper.

This paper is organized as follows.  The geometric theory is
summarized in Section~\ref{sec:geometric} with
Section~\ref{subsec:SU(4)} presenting a review of the way we decompose
$SU(4)$ to separate the purely local operations from the ones which
entangle two qubits, and Section~\ref{subsec:LI} reintroducing the set
of easily-computable numbers which are invariant under the local
operations.  Section~\ref{sec:fctnal} describes the subspace of the
entangling gates which are PEs and introduces the functional that
indicates when we have realized a PE.  The reachable set of PEs for a
generic two-qubit Hamiltonian is discussed in Section~\ref{sec:Weyl}.
Section~\ref{sec:concl} concludes.

\section{Review of the geometric theory for two-qubit gates}
\label{sec:geometric}
\subsection{Decomposition and Parametrization of $SU(4)$}
\label{subsec:SU(4)}

All unitary gates operating on two-qubit states are described by a
$4\times 4$ unitary matrix, an element of the compact Lie group
$U(4)$.  Any such matrix may be written as an element of $SU(4)$
multiplied by a number of modulus $1$, so the sixteen parameters we
use to specify any gate are the phase of this $U(1)$-prefactor (an
angle modulo $\pi/2$) and the fifteen real parameters of $SU(4)$.

Which fifteen parameters we choose are largely up to us; the ones we
use in this work are those arising from the Cartan decomposition of
the Lie algebra of the group, cf. Ref.~\cite{Helgason}.  This
decomposition allows us to write any element of $SU(4)$ as a
combination of two matrices in $SU(2)\otimes SU(2)$ and one in the
maximal Abelian subgroup $\A=SU(4)/SU(2)\otimes SU(2)$.

The utility of this decomposition is apparent when we realise that, in
the computational basis $\{\ket{00},\ket{01},\ket{10},\ket{11}\}$, any
operation which affects only the first qubit is represented by
$U_1\otimes I$, and one affecting only the second is $I\otimes U_2$,
where $U_1$ and $U_2$ are each $2\times 2$ unitary matrices.  These
{\em local} operations, which act separately and independently on the
two qubits, are therefore described by matrices in $SU(2)\otimes
SU(2)$.  The operations which {\em entangle} the two qubits must then
be entirely determined by the matrices from the Abelian subgroup $\A$.
Gates are therefore denoted by equivalence classes living in $\A$; for
example, [CNOT] is the set of gates which are equal to the CNOT gate
up to local operations.

With all of this in hand, we choose the decomposition of $SU(4)$ such
that our matrices take the form
\begin{equation}
U = k_1 A k_2\,,
\end{equation}
where $k_1$ and $k_2$ are $4\times 4$ matrices in $SU(2)\otimes SU(2)$
and $A$ is in the maximal Abelian subgroup $\A$.  Twelve of the
fifteen coordinates necessary to specify any $SU(4)$ element are
included in $k_1$ and $k_2$. Since we work with gates in
$SU(4)$ modulo $SU(2)\otimes SU(2)$, we need only use the three
coordinates $c_1$, $c_2$ and $c_3$ which parametrize the matrix $A$
through
\begin{eqnarray}
  A
  & = &
   \exp\left(-\frac{i}{2}\sum_{j=1}^3c_j
   \sigma_j\otimes\sigma_j\right)
  \nonumber \\ &
  = &
  \prod_{j=1}^3\left[I\otimes I\cos\left(\frac{c_j}{2}\right)-
  i\sigma_j\otimes\sigma_j\sin\left(\frac{c_j}{2}\right)\right]\,,
\label{eq:cartan_decomp}
\end{eqnarray}
where $\sigma_{x,y,z}$ are the usual Pauli matrices.  (Later in this
article we shall use the shorthand $\sigma_i^{(1)}=\sigma_i\otimes I$
and $\sigma_i^{(2)}=I\otimes\sigma_i$.)  To ensure that each $U$ is
given by a unique set of coordinates, we must restrict $c_1$, $c_2$
and $c_3$ to the Weyl chamber $W$ given by
\begin{eqnarray*}
&&0\leq c_3\leq c_2\leq c_1\leq \frac{\pi}{2}
\quad \mbox{or}\\
&&\frac{\pi}{2}<c_1<\pi,\quad 0\leq c_3\leq c_2<\pi-c_1\,,
\end{eqnarray*}
i.e., within the tetrahedron whose vertices are at $(0,0,0)$,
$(\pi,0,0)$, $(\pi/2,\pi/2,0)$ and
$(\pi/2,\pi/2,\pi/2)$~\cite{ZhangPRA03}.

\subsection{Local Invariants}
\label{subsec:LI}
Although $c_1$, $c_2$ and $c_3$ are defined in a straightforward
manner, actually determining their values for a general element of
$SU(4)$ can be difficult.  Fortunately, there are three alternative
parameters which can be used as coordinates for local equivalence classes
on $\mathcal{A}$ which are
far easier to obtain.

If we change from the standard computational basis
$\{\ket{00},\ket{01},\ket{10},\ket{11}\}$ to a Bell basis given by
\begin{eqnarray*}
  \Bigg\{
  \frac{1}{\sqrt{2}} \left(  \ket{00} - i \ket{11}\right)&,&
  - \frac{1}{\sqrt{2}} \left(i \ket{01} -   \ket{10}\right),\\ \;
  - \frac{1}{\sqrt{2}} \left(i \ket{01} +   \ket{10}\right)&,&
  + \frac{1}{\sqrt{2}} \left(  \ket{00} + i \ket{11}\right)
  \Bigg\},
\end{eqnarray*}
then our $SU(4)$ matrices become $U_{\mathrm{B}}=Q^{\dagger}UQ=
Q^{\dagger}k_1Ak_2Q$, where
\begin{eqnarray*}
Q&=&\frac{1}{\sqrt{2}}\left(\begin{array}{cccc}
1&0&0&i\\0&i&1&0\\0&i&-1&0\\1&0&0&-i
\end{array}\right).
\end{eqnarray*}
The eigenvalues of the matrix $m=U_{\mathrm{B}}^{\mathrm{T}}U_{\mathrm{B}}$
determine the local invariants of $U$~\cite{Makhlin}.  The
characteristic equation of $m$ is
\begin{eqnarray*}
\lambda^4-\tr(m)\lambda^3+\frac{1}{2}\left[\tr^2(m)-\tr\left(m^2\right)
  \right]\lambda^2 -\tr^*(m)\lambda+1&=&0,
\end{eqnarray*}
and so $\tr(m)$ and $\tr(m^2)$ give the local invariants.  These are
complex numbers. Instead we may take as local invariants the three
real numbers
\begin{eqnarray*}
g_1&=&\frac{1}{16}\mathrm{Re}\left\{\tr^2(m)\right\}\,,\,
g_2=\frac{1}{16}\mathrm{Im}\left\{\tr^2(m)\right\}\,,\\
g_3&=&\frac{1}{4}\left[\tr^2(m)-\tr\left(m^2\right)\right]\,.
\end{eqnarray*}
Since $m$, $m^2$ and their traces are readily computable using the
simplest of matrix operations, values for $g_1$, $g_2$ and $g_3$ can
be easily obtained for any $U\in SU(4)$.

\begin{figure*}[tb]
\centering
\includegraphics[width=0.45\linewidth]{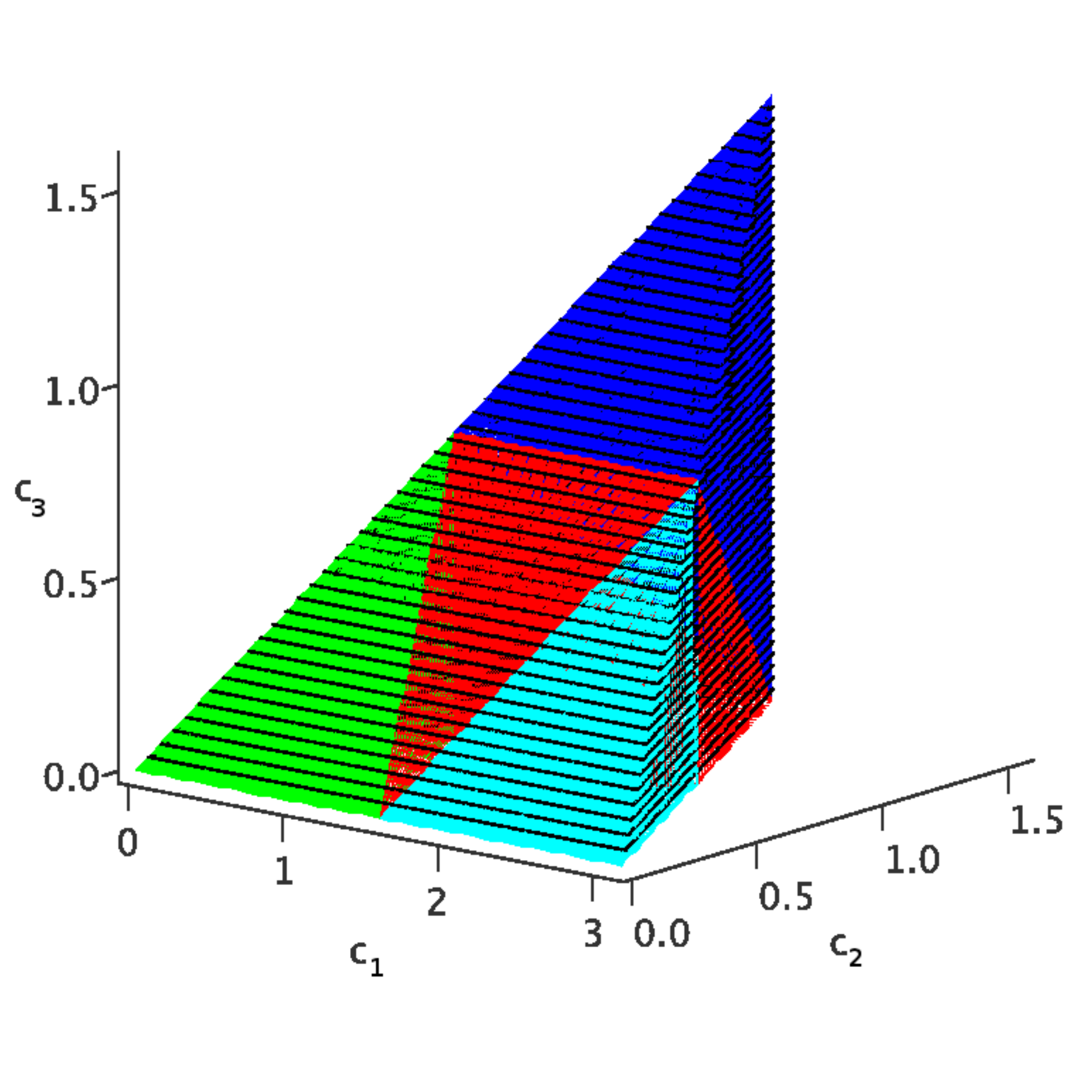}\hfill
\includegraphics[width=0.45\linewidth]{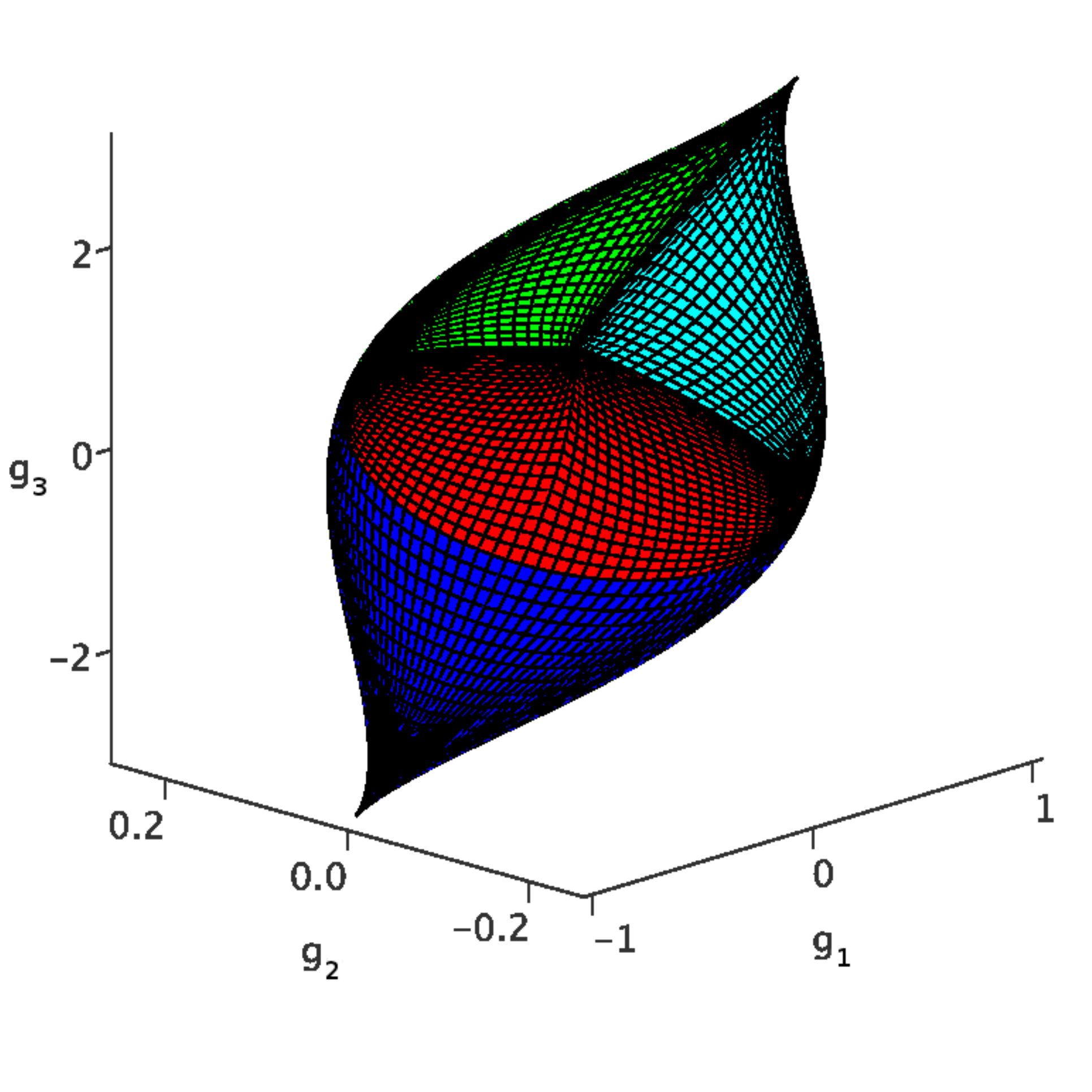}
\caption{(Color online) The Weyl chamber in $c_1c_2c_3$ space (left)
  and its embedding in $g_1g_2g_3$ space (right).  In both, $W_0$ is
  in green, $W_0^*$ in cyan, $W_1$ in blue and $W_{\mathrm{PE}}$ in
  red.  (The contours shown are purely for illustrative purpose.)}
\label{fig:full-Weyl}
\end{figure*}

\begin{figure*}[tb]
\centering
\includegraphics[width=0.45\linewidth]{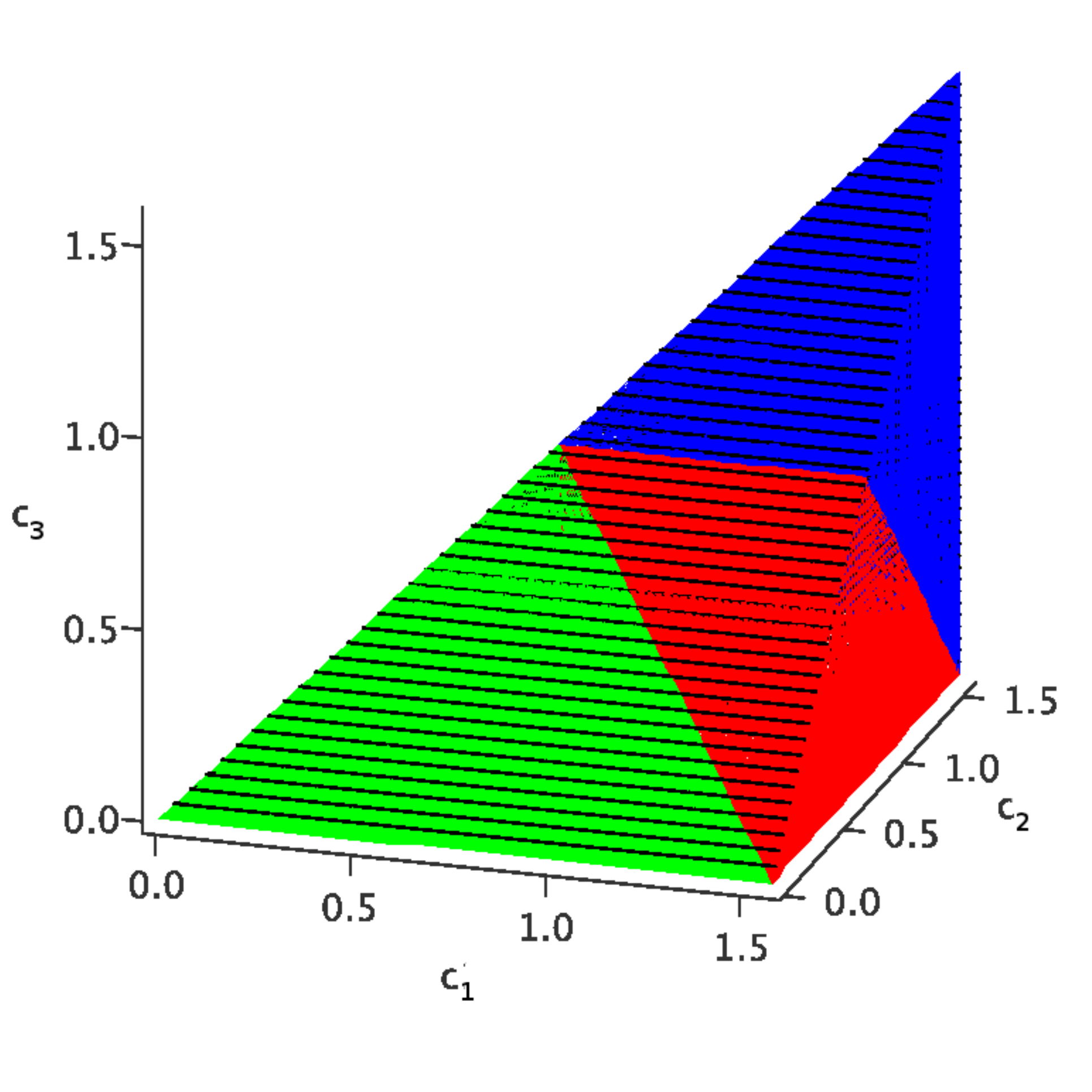}\hfill
\includegraphics[width=0.45\linewidth]{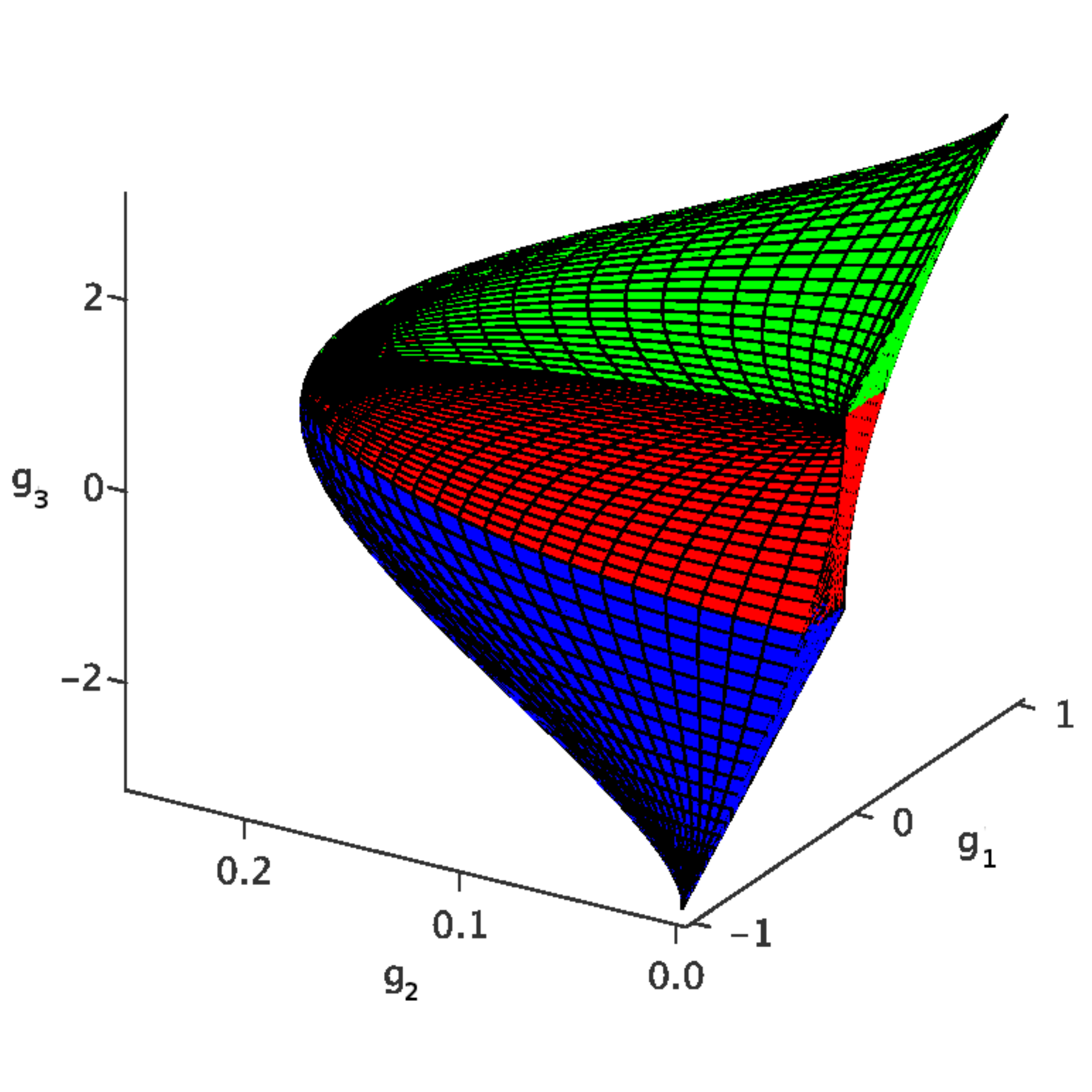}
\caption{(Color online) The $0\leq c_1\leq\pi/2$ half of the Weyl
  chamber in $c_1c_2c_3$ space (left) and the corresponding $g_2\geq
  0$ half in $g_1g_2g_3$ space (right).  In both cases, the full
  chamber is obtained by reflection across the cross-section at the
  right of each figure.}
\label{fig:half-Weyl}
\end{figure*}

Since $g_1$, $g_2$, $g_3$ are local invariants, they must be functions
only of $c_1$,
$c_2$ and $c_3$; some computation shows that they are, and have the
explicit forms
\begin{eqnarray*}
g_1&=&\frac{1}{4}\big[\cos\left(2c_1\right)+\cos\left(2c_2\right)
  +\cos\left(2c_3\right)\\
  &&+\cos\left(2c_1\right)\cos\left(2c_2\right)
  \cos\left(2c_3\right)\big]\,,\\
g_2&=&\frac{1}{4}\sin\left(2c_1\right)\sin\left(2c_2\right)
\sin\left(2c_3\right) \,,\\
g_3&=&\cos\left(2c_1\right)+\cos\left(2c_2\right)+\cos\left(2c_3\right)\,.
\end{eqnarray*}
These can be used to embed the tetrahedron defining the Weyl chamber
into $g_1g_2g_3$-space; both spaces are shown in
Figure~\ref{fig:full-Weyl}, with cross-sections shown in
Figure~\ref{fig:half-Weyl}.  The coordinates for the labeled points
in both spaces are given in Table~\ref{tab:points}.
\begin{table}[tb]
\centering
\begin{tabular}{c|ccc|ccc}
\hline
point (gate)&$c_1$&$c_2$&$c_3$&$g_1$&$g_2$&$g_3$\\
\hline\hline
$O$, $A_1$ ([$\openone$]) & $0$, $\pi$&$0$&$0$&$1$&$0$&$3$\\
$A_2$ ([DCNOT])&$\pi/2$&$\pi/2$&$0$&$0$&$0$&$-1$\\
$A_3$ ([SWAP])&$\pi/2$&$\pi/2$&$\pi/2$&$-1$&$0$&$-3$\\
$B$ ([B-Gate])&$\pi/2$&$\pi/4$&$0$&$0$&$0$&$0$\\
$L$ ([CNOT])&$\pi/2$&$0$&$0$&$0$&$0$&$1$\\
$P$ ([$\sqrt{\mbox{SWAP}}$])&$\pi/4$&$\pi/4$&$\pi/4$&$0$&$1/4$&$0$\\
$Q$, $M$&$\pi/4$, $3\pi/4$&$\pi/4$&$0$&$1/4$&$0$&$1$\\
$N$&$3\pi/4$&$\pi/4$&$\pi/4$&$0$&$-1/4$&$0$\\
$R$&$\pi/2$&$\pi/4$&$\pi/4$&$-1/4$&$0$&$-1$\\
\hline
\end{tabular}
\caption{The coordinates of selected points in the Weyl chamber (see
  Figures~\ref{fig:full-Weyl} and~\ref{fig:half-Weyl}) in $c_1c_2c_3$
  space and their corresponding local invariants.}\label{tab:points}
\end{table}

A particular combination  which is quite useful is
$\sqrt{g_1^2+g_2^2}$; a quick calculation shows that
\begin{eqnarray*}
g_1^2+g_2^2&=&\frac{1}{16}\big[1+\cos\left(2c_1\right)
\cos\left(2c_2\right)\\&&+\cos\left(2c_1\right)\cos\left(2c_3\right)
+\cos\left(2c_2\right)\cos\left(2c_3\right)\big]^2.
\end{eqnarray*}
It is straightforward to confirm that the quantity inside the square
brackets is always non-negative inside the Weyl chamber, so
\begin{eqnarray*}
\sqrt{g_1^2+g_2^2}&=&\frac{1}{4}\big[1+\cos\left(2c_1\right)
  \cos\left(2c_2\right)\\
  &&+\cos\left(2c_1\right)\cos\left(2c_3\right)
  +\cos\left(2c_2\right)\cos\left(2c_3\right)\big].
\end{eqnarray*}

\section{A functional for perfect entanglers}
\label{sec:fctnal}

The elements of $SU(4)$ which perfectly entangle two-qubit states all
lie within the subset of the Weyl chamber $W$ bounded by the planes
$c_1+c_2=\pi/2$, $c_1-c_2=\pi/2$ and $c_2+c_3=\pi/2$.  This region is
the 7-faced polyhedron with vertices at $(\pi/2,0,0)$,
$(\pi/4,\pi/4,0)$, $(3\pi/4,\pi/4,0)$, $(\pi/2,\pi/2,0)$,
$(\pi/4,\pi/4,\pi/4)$ and $(3\pi/4,\pi/4,\pi/4)$~\cite{ZhangPRA03}.
$W$ is thus divided up into four regions:
\begin{enumerate}
\item $W_{\mathrm{PE}}$, the perfect entanglers themselves.
\item $W_0$, the region between the origin (i.e., the identity
  element) and $W_{\mathrm{PE}}$, the tetrahedron bounded by (but not
  including) the wall $c_1+c_2=\pi/2$.  All three local invariants are
  positive in this region.
\item $W_0^*$, between $(\pi,0,0)$ and $W_{\mathrm{PE}}$, bounded by
  $c_1-c_2=\pi/2$.  In this region, $g_1$ and $g_3$ are positive and
  $g_2$ is negative.  In fact, $W_0^*$ can be obtained from $W_0$ via
  the transformation $(g_1,g_2,g_3)\rightarrow (g_1,-g_2,g_3)$.
\item $W_1$, between $W_{\mathrm{PE}}$ and the [SWAP] gate at
  $(\pi/2,\pi/4,\pi/4)$, bounded by $c_2+c_3=\pi/2$.  $g_1$ and $g_3$
  are both negative and $g_2$ can have any sign.
\end{enumerate}
One can construct functions based on a parametrization of $W_{PE}$
either in terms of $(c_1,c_2,c_3)$ or in terms of
$(g_1,g_2,g_3)$. In the following, we will refer to $(c_1,c_2,c_3)$ as
the Weyl coordinates and to $(g_1,g_2,g_3)$ as the local invariants or
Makhlin coordinates.

\subsection{Gate fidelity for  perfect entanglers in terms of
  the Weyl coordinates $c_1$, $c_2$, $c_3$}

In order to define a fidelity for an arbitrary perfect entangler in
terms of the Weyl coordinates $c_1$, $c_2$, $c_3$, we generalize the
notion of the gate fidelity for a specific desired gate $V$,
\begin{eqnarray*}
 F&=&\frac{1}{4}\left|\tr\left( U^\dagger V\right)\right|\,,
\end{eqnarray*}
where $U$ is the actually-implemented gate, and we assume $U\in
SU(4)$.  Allowing for complete freedom in the local transformations,
this becomes
\begin{eqnarray*}
  F&=&\max_{k_1,k_2 \in SU(2)\otimes
    SU(2)}\frac{1}{4}\mathrm{Re}\left\{\tr\left(
        U^\dagger k_1 V k_2\right)\right\}\,,
\end{eqnarray*}
where we have substituted the modulus by the real part implying that
without loss of generality we can choose the global phase of the local
transformations such that the trace is real.  The maximum over all
local transformations $k_1$, $k_2$ is difficult to evaluate.  However,
the local transformations can be chosen such that $U$ and $V$ are
given by their canonical forms $A_U=\exp[-i/2\sum_j  c_j^U \sigma_j\sigma_j]$
and $A_V=\exp[-i/2\sum_j  c_j^V \sigma_j\sigma_j]$. We denote this choice
by $k_i=k_{i,U}k_{i,V}$.  It can be shown that the partial derivatives
of $F$ with respect to the $k_i$ vanish and that $F=1$ for $U=V$. The
latter simply follows from equality of the Weyl coordinates. The
partial derivatives 
are obtained by parametrizing the $k_i$ as elements of $SU(2)\otimes
SU(2)$ and the canonical forms of the non-local parts by $c_1$, $c_2$,
$c_3$. This choice of the local transformations yields
\begin{eqnarray*}
  F&=&\frac{1}{4}\mathrm{Re}\left\{\tr
  \left( U^\dagger k_{1,U}k_{1,V}^\dagger Vk_{2,V}^\dagger k_{2,U}\right)\right\}\\
  &=&\frac{1}{4}\mathrm{Re}\left\{\tr\left( A_U^\dagger  A_V\right)\right\}\\
  &=&\frac{1}{4}\mathrm{Re}\left\{ \tr\left(
      Q^\dagger A_U^\dagger Q Q^\dagger A_VQ\right)\right\}\\
  &=&\frac{1}{4}\mathrm{Re}\left\{ \tr\left( F_U^\dagger F_V \right) \right\}
\end{eqnarray*}
with
\begin{eqnarray*}
  F_U&=&Q^\dagger A_{U} Q\\
  &=&\mathrm{diag}(\mathrm{e}^{\imath\frac{c_{1}-c_{2}+c_{3}}{2}},
  \mathrm{e}^{\imath\frac{c_{1}+c_{2}-c_{3}}{2}},
  \mathrm{e}^{\imath\frac{-c_{1}-c_{2}-c_{3}}{2}},
  \mathrm{e}^{\imath\frac{-c_{1}+c_{2}+c_{3}}{2}})\\
  &=&\mathrm{diag}(\mathrm{e}^{\imath\phi_{1,U}},
  \mathrm{e}^{\imath\phi_{2,U}},\mathrm{e}^{\imath\phi_{3,U}},
\mathrm{e}^{\imath\phi_{4,U}})\,
\end{eqnarray*}
and $F_V=Q^\dagger A_{V} Q$, respectively.
Inserting the explicit forms of $F_U$ and $F_V$, we obtain
\begin{eqnarray}
  F &=&\frac{1}{4}\mathrm{Re}\left\{\tr\left(F_U^\dagger F_V\right)\right\}
\nonumber\\
  &=& \frac{1}{4}\sum_{j=1}^4
  \cos(\varphi_{j,U}-\varphi_{j,V})\nonumber \\
  &=&\frac{1}{4}\bigg(\cos\frac{\Delta c_1 -\Delta c_2 +
\Delta c_3}{2} +\cos\frac{\Delta c_1 +\Delta c_2 - \Delta c_3}{2}
\nonumber\\
  &&+\cos\frac{\Delta c_1 +\Delta c_2 + \Delta c_3}{2}
  +\cos\frac{\Delta c_1 -\Delta c_2 - \Delta c_3}{2}
\bigg)\nonumber\\
  &=&\cos\frac{\Delta c_1}{2}\cos\frac{\Delta c_2}{2}\cos\frac{\Delta
    c_3}{2}\approx 1-\frac{|\Delta
    \vec{c}|^2}{8}\label{eq:fidelity-nonlocal-unitary}\,,
\end{eqnarray}
where $\Delta c_i=c_{U,i}-c_{V,i}$. In order to find the closest
perfect entangler $V$ for a given gate $U$, we have to maximize the
fidelity given by Eq.~\eqref{eq:fidelity-nonlocal-unitary} with
respect to $c_{V,i}$.  To this end, we can exploit that the sectors
$W_0$, $W_0^*$, $W_1$ are separated from the polyhedron $W_{PE}$ by
three planes, and $U$ is a perfect entangler if and only if
\begin{eqnarray*}
 c_1+c_2\geq \frac\pi 2\mbox{, }
 c_1-c_2\leq \frac\pi 2&\mbox{and}& c_2+c_3\leq \frac\pi 2\,.
\end{eqnarray*}
If $U$ lies in the polyhedron of perfect entanglers we can simply
choose $V=U$ and arrive at perfect fidelity $F=1$.  If $U\in W_0$, we
have $c_1+c_2\leq \frac\pi 2$, and the closest perfect entangler both
in terms of fidelity and distance of the Weyl coordinates
is given by the projection
of $U$ onto the wall, i.e., $c_{V,1}=\frac{\pi}{4}+\frac{c_{U,1} -
  c_{U,2}}{2}$, $c_{V,2}= \frac{\pi}{4}+\frac{c_{U,2} - c_{U,1}}{2}$,
and $c_{V,3}=c_{U,3}$.  The distance vector between $U$ and $V$ as a
function of the Weyl coordinates is then given by
\begin{eqnarray*}
\Delta\vec{c}&=& \left(
  \frac{c_{U,1} + c_{U,2}}{2}- \frac{\pi}{4},
  \frac{c_{U,1} + c_{U,2}}{2}- \frac{\pi}{4}, 0\right)\,.
\end{eqnarray*}
With the analogous approach for $W_0^*$ and $W_1$ and using
Eq.~\eqref{eq:fidelity-nonlocal-unitary}, we arrive at
\begin{eqnarray*}
F_{PE}(U)=\begin{cases}
\cos^2\frac{c_{U,1}+c_{U,2}-\frac{\pi}{2}}{4}\,,\qquad c_1+c_2\leq\frac{\pi}{2}\\
\cos^2\frac{c_{U,2}+c_{U,3}-\frac{\pi}{2}}{4}\,,\qquad c_2+c_3\geq\frac{\pi}{2}\\
\cos^2\frac{c_{U,1}-c_{U,2}-\frac{\pi}{2}}{4}\,,\qquad c_1-c_2\geq\frac{\pi}{2}\\
1\qquad\qquad\text{otherwise (inside $W_{PE}$).}
\end{cases}
\end{eqnarray*}
As desired, this fidelity is a function of $c_{U,i}$; it equals one if
and only if $U$ is a perfect entangler and is smaller than 1
otherwise.  $F_{PE}(U)$ can be used for optimization if no analytic
gradients with respect to the states are needed.

Often the dynamics may explore a Hilbert space that is larger than the
logical subspace of the qubits. The evolution in the logical
subspace may then correspond to a non-unitary gate $\tilde U$. Employing a
singular value decomposition of $\tilde U$ and renormalizing the
singular values, a unitary approximation $U$ of $\tilde U$ is obtained
analogously to the unitary case. This
allows to utilize the same ideas that have lead to the fidelity
$F_{PE}$ defined above. The gate fidelity $F$ becomes
\begin{eqnarray*}
  F&=&\frac{1}{4}\left|\tr\left(\tilde{U}^\dagger V\right)\right|  \,,
\end{eqnarray*}
where $V=k_{1,U} A_V k_{2,U}$ and $A_V$ is the canonical form of the
perfect entangler closest to the unitary approximation $U$, as
measured by the distance in Weyl coordinates. In order
to avoid explicit calculation of the $k_{i,U}$ (which would have to be
done in every iteration step of an optimization algorithm), we find
the lower bound on the fidelity,
\begin{eqnarray*}
 F&=&\frac{1}{4}\left|\tr\left(\tilde{U}^\dagger V\right)\right|
 =\frac{1}{4}\mathrm{Re}\left\{\tr\left(\tilde U^\dagger V\right)\right\}\\
 &=&\frac{1}{4}\mathrm{Re}\left\{\tr\left(U^\dagger V\right)\right\}+
 \frac{1}{4}\mathrm{Re}\left\{\tr\left((\tilde U -U)^\dagger V\right)\right\}\\
 &\geq& \frac{1}{4}\mathrm{Re}\left\{\tr\left(U^\dagger V \right)\right\}
 - \left|\frac{1}{4}\tr\left((\tilde U -U)^\dagger V\right)\right|\\
 &\geq& F_{PE}(U) - ||\tilde U - U||\,,
\end{eqnarray*}
where we have first used the choice of $V$ that makes the trace real,
and then used both the Cauchy-Schwarz inequality and $||V||=1$.

\subsection{Perfect entanglers and the local invariants}

For optimization algorithms that utilize gradient information it is
necessary to express the functional in a way that allows for analytic
expressions of the derivatives~\cite{ReichJCP12}.  This is not the
case if the functional is expressed in terms of the Weyl coordinates
$(c_1,c_2,c_3)$~\cite{MuellerPRA11}.  We therefore seek to express the
boundaries of the polyhedron $W_{PE}$ in terms of the local
invariants $(g_1,g_2,g_3)$.

Let us first look at the boundary with $W_0$: it is defined by the plane
$c_1+c_2=\pi/2$, and along this wall, $\cos(2c_2)=-\cos(2c_1)$ and
$\sin(2c_2)=\sin(2c_1)$.  This means that the values of the local
invariants on this wall depend only on $c_1$ and $c_3$ through
\begin{eqnarray*}
g_1&=&\frac{1}{4}\sin^2\left(2c_1\right)\cos\left(2c_3\right),\\
\sqrt{g_1^2+g_2^2}&=&\frac{1}{4}\sin^2\left(2c_1\right),\\
g_3&=&\cos\left(2c_3\right).
\end{eqnarray*}
We can eliminate $c_1$ and $c_3$ entirely from the above to give
\begin{eqnarray*}
g_3&=&\left\{
\begin{array}{cl}
\frac{g_1}{\sqrt{g_1^2+g_2^2}}&g_1\neq 0\mbox{ or }g_2\neq 0,\\
1&g_1=g_2=0
\end{array}\right.
\end{eqnarray*}
as the equation defining the PE boundary in terms of the local invariants.
If we repeat this analysis for the walls separating $W_0^*$ and $W_1$
from $W_{\mathrm{PE}}$, we find that the {\em same} equation describes
them all.  So any $U$ lying precisely on the
boundary of $W_{\mathrm{PE}}$ has local invariants satisfying
$g_3=g_1/\sqrt{g_1^2+g_2^2}$.

This suggests the definition of a function $d$ which depends
on an $SU(4)$ matrix $U$ via its local invariants and vanishes on the
boundary of $W_{\mathrm{PE}}$:
\begin{eqnarray}
  \label{eq:PEfunction}
  d\left(g_1,g_2,g_3\right)&=&g_3\sqrt{g_1^2+g_2^2}-g_1\,.
\end{eqnarray}
This is not the only combination of the local invariants which
vanishes on the boundary of $W_{\mathrm{PE}}$; the reason we choose
this particular definition of $d$ comes from the fact that
it is continuous for all values of $g_1$, $g_2$ and $g_3$. When we
rewrite it in terms of the Weyl coordinates,
we obtain the particularly simple form
\begin{eqnarray*}
d&=&\frac{1}{4}\left[\cos\left(2c_1\right)+\cos\left(
  2c_2\right)\right]\left[\cos\left(2c_1\right)+\cos\left(2c_3\right)\right]\\
&&\times\left[\cos\left(2c_2\right)+\cos\left(2c_3\right)\right].
\end{eqnarray*}
It is this form which allows us to see immediately that $d$ is
manifestly positive in $W_0$; thus, in terms of the local invariants,
all points in $W_0$ satisfy $g_3\sqrt{g_1^2+g_2^2}-g_1>0$.
We noted above that $W_0^*$ is simply the mirror-reflection of $W_0$,
since we may obtain it by changing the sign of $g_2$; thus, in
reality, $W_0$ and $W_0^*$ are not disconnected in terms of the local
invariants, but are joined along the $g_2=0$ plane.  This is seen
explicitly in Figure~\ref{fig:full-Weyl}, where $W_0\cup W_0^*$
consists of the green and cyan regions of the Weyl chamber.

\begin{figure}[tb]
  \centering
  \includegraphics[width=0.9\linewidth]{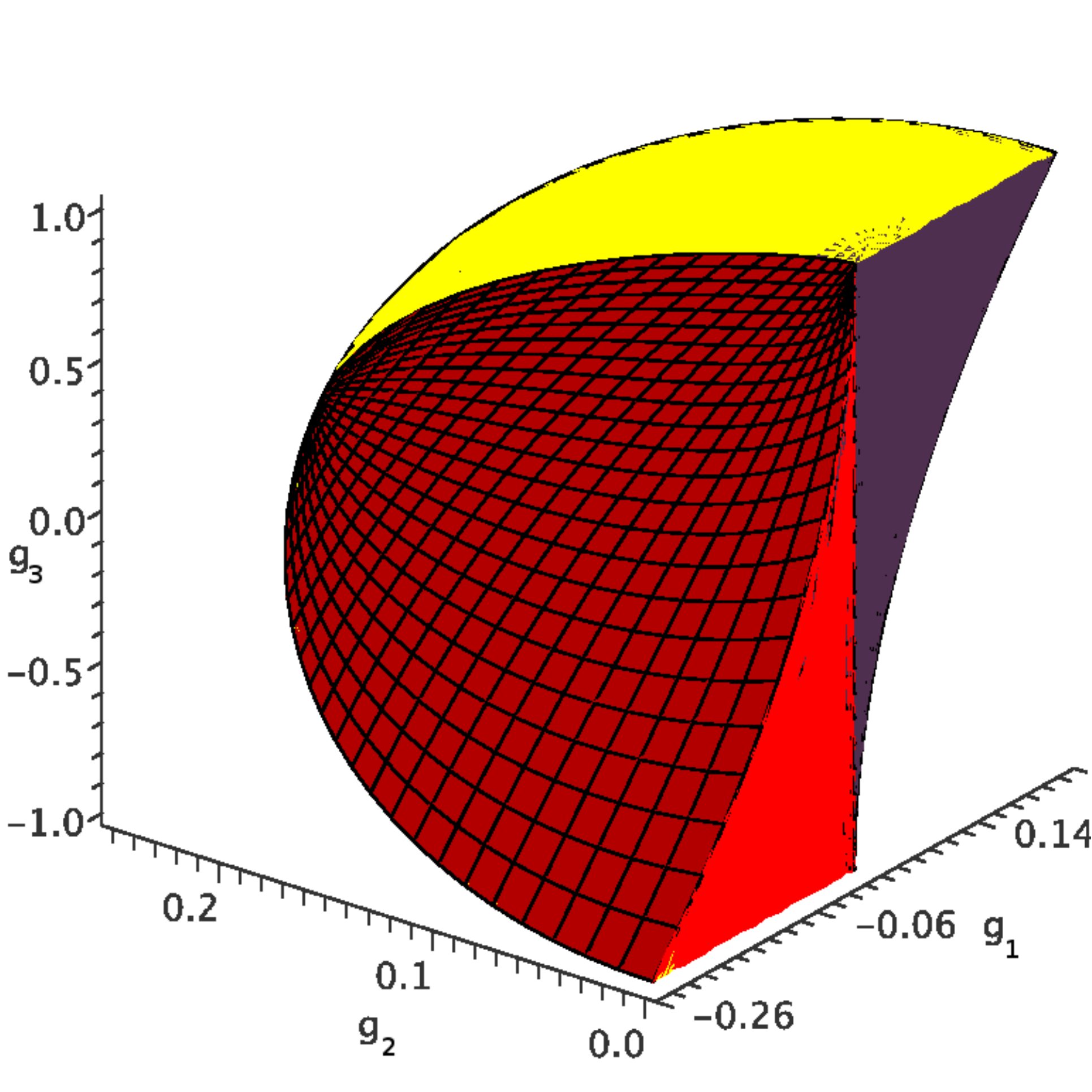}
  \caption{The $g_2\geq 0$ half of the set of perfect entanglers
    $W_{\mathrm{PE}}$ in $g_1g_2g_3$ space.  This space is divided
    into three regions: the red volume, where $d>0$; the violet
    volume, where $d<0$; and the surface composed of the boundary
    between them and the uppermost (yellow) and lowermost (obscured)
    surfaces, where $d=0$.  }
  \label{fig:PE-wall}
\end{figure}

In $g$-space, the boundary separating $W_0\cup W_0^*$ from
$W_{\mathrm{PE}}$ is a single continuous surface.  To be precise, if
we use cylindrical coordinates $(\rho,\phi,z)$ defined by
$g_1=\rho\cos\phi$, $g_2=\rho\sin\phi$ and $g_3=z$, the boundary is
given by the surface
\begin{eqnarray*}
z=\cos\phi&\mbox{with}&-\frac{\pi}{2}\leq\phi\leq\frac{\pi}{2},\,\frac{1}{4}
\sin^2\phi\leq\rho\leq\frac{1}{4}.
\end{eqnarray*}
The part of this wall adjoining $W_0$ is the yellow surface
illustrated in Figure~\ref{fig:PE-wall}.
As a result, if optimization starts from a gate $U$ in $W_0\cup
W_0^*$, then $d(g_1,g_2,g_3)=g_3\sqrt{g_1^2+g_2^2}-g_1$ is an
optimization function to reach a PE gate: we know that $d>0$ for the
initial gate and it reaches zero at the boundary with
$W_{\mathrm{PE}}$.  However, $d$ vanishes elsewhere as well: not only
on the boundary between $W_{\mathrm{PE}}$ and $W_1$, but everywhere on
the surface $z=\cos\phi$.  This surface is comprised not only of the
boundaries that $W_{\mathrm{PE}}$ has with $W_1$ and $W_0\cup W_0^*$
but also the boundary between the red and violet regions in
Figure~\ref{fig:PE-wall}.  However, this surface lies entirely within
$W_{\mathrm{PE}}$, so the {\em only} gates $U$ for which
$d(g_1,g_2,g_3)$ vanishes are perfect entanglers.


However, $d$ alone cannot tell us if we continue into the interior of
$W_{\mathrm{PE}}$.  If $U$ happens to cross the curve $z=\cos\phi$,
$\rho=\sin^2\phi/4$, then either $d(g_1,g_2,g_3)$ becomes positive and
we have a PE, or it becomes negative and we are in $W_1$ and do not
have a PE.  In either of these two cases, the value of $d$ alone will
not be a good enough indicator of whether we have evolved to a PE; further
information might be necessary.


\subsection{An optimization functional for perfect entanglers}

The discussion of the previous two sections motivates our formulation
of a functional $\mathcal{D}(U)$ that provides a {\em definitive}
answer as to whether or not an $SU(4)$ gate $U$ is locally equivalent
to a perfect entangler.  That is, the functional vanishes if $U$ is a
perfect entangler and is positive otherwise.


The functional $\mathcal{D}(U)$ is based on the function
$d(g_1,g_2,g_3)$ but also takes into account in which sector of the
Weyl chamber -- $W_0$, $W_0^*$, $W_1$ or $W_{\mathrm{PE}}$ -- the
local equivalence class of the gate $U$ is located.  Its construction
is presented below:
\begin{enumerate}
\item Compute the three Makhlin invariants $g_1$, $g_2$ and $g_3$ for
  $U$ as usual.
\item Next, find the three roots $z_1$, $z_2$ and $z_3$ of the cubic
  equation
\begin{equation*}
~~~~~~z^3-g_3z^2+\left(4\sqrt{g_1^2+g_2^2}-1\right)z+\left(g_3-4g_1\right)=0
\end{equation*}
ordered such that $-1\leq z_1\leq z_2\leq z_3\leq 1$.  These roots --
which are functions of $g_1$, $g_2$ and $g_3$ -- facilitate the
inverse map $\left(g_1,g_2,g_3\right) \rightarrow
\left(c_1,c_2,c_3\right)$ and thus provide the location of the gate
within the $c$-space Weyl chamber~\cite{WattsEnt13}.\\
\item Define $d$ as in Eq.~\eqref{eq:PEfunction} and $s$ as
\begin{eqnarray*}
s\left(g_1,g_2,g_3\right)&:=&\pi-\cos^{-1}z_1-\cos^{-1}z_3
\end{eqnarray*}
The definition of the functional $\mathcal{D}$ depends on the signs of
these two functions:
\begin{enumerate}
\item If $d$ and $s$ are both positive, then
\begin{eqnarray*}
\mathcal{D}(U)&=&g_3\sqrt{g_1^2+g_2^2}-g_1.
\end{eqnarray*}
\item If $d$ and $s$ are both negative, then
\begin{eqnarray*}
\mathcal{D}(U)&=&g_1-g_3\sqrt{g_1^2+g_2^2}.
\end{eqnarray*}
\item In any other case,
\begin{eqnarray*}
\mathcal{D}(U)&=&0.
\end{eqnarray*}
\end{enumerate}
\end{enumerate}
This gives the desired functional, one that is zero when the two-qubit
gate is a perfect entangler and positive otherwise.  Its evaluation
requires only the Makhlin invariants and a way of finding the largest
and smallest roots of a cubic equation.  The functional is also
differentiable and straightforward to implement within the framework of
optimal control.

\section{Controllability in the Weyl chamber}
\label{sec:Weyl}

Optimization towards an arbitrary perfect entangler is most meaningful
if the system dynamics allows the polyhedron of perfect entanglers to
be approached from more than one direction or, more generally, for
optimization paths in the Weyl chamber that explore more than one
dimension.  We therefore investigate the corresponding requirements on
a generic two-qubit Hamiltonian,
\begin{eqnarray}
  H[u_{1}(t), u_{2}(t)]
  &=&  \sum_{\alpha=1,2} \frac{\omega_{\alpha}}{2} \sigma_z^{(\alpha)}
      + u_1(t) \left( \sigma_x^{(1)} + \lambda \sigma_x^{(2)}
      \right)\nonumber \\
  && + u_2(t) \left(\sigma_x^{(1)} \sigma_x^{(2)}
         +\sigma_y^{(1)} \sigma_y^{(2)}\right)\,.
  \label{eq:fullH}
\end{eqnarray}
Here, $\sigma_{i}^{(\alpha)}$ is the $i^{\mathrm{th}}$ Pauli operator
acting on the $\alpha^{\mathrm{th}}$ qubit of transition frequency
$\omega_{\alpha}$, $u_1(t)$ the
single-qubit control field, where $\lambda$ describes how strongly $u_1(t)$
couples to the second qubit relative to the first one, and $u_2(t)$ is the
two-qubit interaction control field.
As discussed in the sequel to this paper, Eq.~\eqref{eq:fullH} is
used to model qubits realized with superconducting circuits.

We analyze the solutions to the differential equation
\begin{equation}
\dot{U}\left(t\right)=-iH\left[u\left(t\right)\right]U\left(t\right),
\quad U\left(0\right)=\openone\,
\label{eq:U_dgl}
\end{equation}
for the unitary transformations $U$ generated by the
Hamiltonian~\eqref{eq:fullH}.  The reachable set of unitary
transformations for a Hamiltonian is given in terms of the
corresponding dynamical Lie algebra.  It can be generated by taking
the terms in \eqref{eq:fullH} as a basis (neglecting
orthonormalization for simplicity),
\begin{eqnarray*}
  \sigma_{z}^{(1)} \,,\,
  \sigma_{z}^{(2)} \,,\,
  \sigma_x^{(1)}  + \lambda \sigma_x^{(2)} \,,\,
  \sigma_x^{(1)} \sigma_x^{(2)} + \sigma_y^{(1)} \sigma_y^{(2)}\,,
\end{eqnarray*}
and constructing the repeated Lie brackets of these operators. This quickly
yields all 15 canonical basis operators of $SU(4)$, consisting of the
single-qubit operators
  $\sigma_{x}^{(1)}$,
  $\sigma_{x}^{(2)}$,
  $\sigma_{y}^{(1)}$,
  $\sigma_{y}^{(2)}$,
  $\sigma_{z}^{(1)}$, and
  $\sigma_{z}^{(2)}$,
as well as the entangling operators
  $\sigma_{x}^{(1)}\sigma_{y}^{(2)}$,
  $\sigma_{y}^{(1)}\sigma_{x}^{(2)}$,
  $\sigma_{y}^{(1)}\sigma_{z}^{(2)}$,
  $\sigma_{z}^{(1)}\sigma_{y}^{(2)}$,
  $\sigma_{x}^{(1)}\sigma_{z}^{(2)}$,
  $\sigma_{z}^{(1)}\sigma_{x}^{(2)}$,
  $\sigma_{x}^{(1)}\sigma_{x}^{(2)}$,
  $\sigma_{y}^{(1)}\sigma_{y}^{(2)}$, and
  $\sigma_{z}^{(1)}\sigma_{z}^{(2)}$.
Hence the system is completely controllable, and any point in the Weyl
chamber can be reached.

\begin{figure*}[tb]
  \centering
  \includegraphics{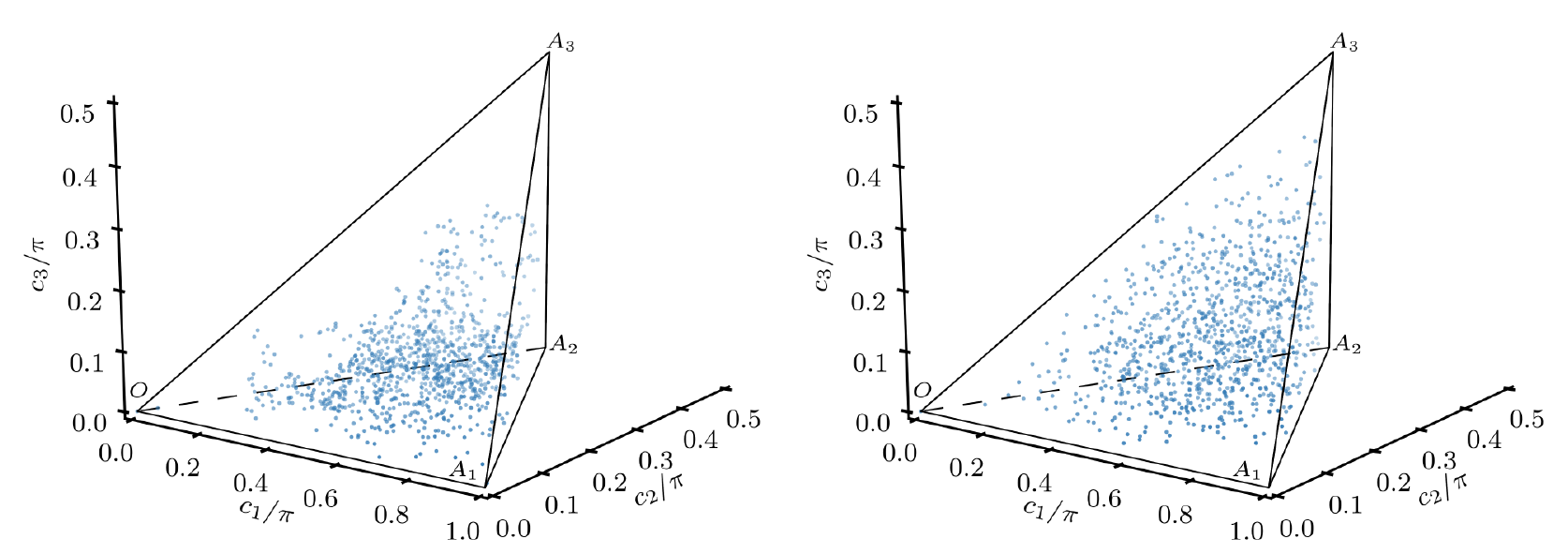}
  \caption{(Color online) Sampling of reachable points in the Weyl
    chamber, obtained by solving Eq.~\eqref{eq:U_dgl}
    for  the Hamiltonian~\eqref{eq:fullH}
    ($\lambda=1$), a random pulse $u_{1}(t) \in [0,1]$,
    constant $u_{2}(t) \equiv 10^{-3}$, and 1000 time steps.
    On the left, result for $\omega_1 = 1.0 \neq
    \omega_2=1.1$, providing the full set of 15 generators in the Lie
    algebra.
  On the right, result for $\omega_1 = \omega_2=1$, providing
  9 generators. In both cases, every point in the Weyl chamber can be reached.}
  \label{fig:fullControl}
\end{figure*}
The complete controllability can be verified numerically, by solving
Eq.~\eqref{eq:U_dgl} for a random sequence of pulse values. The resulting gates
are shown in the left of Fig.~\ref{fig:fullControl}, and demonstrate full
controllability, since there are points in all regions of the Weyl chamber.
Continuing the procedure to infinity would eventually fill the entire chamber.
Neither setting $u_2(t)$ constant nor choosing $\lambda=0$ places any
restrictions on the controllability -- indeed it is sufficient if either
the single qubit terms or the interaction term is controllable.
While the controllability in this example was analyzed for
arbitrary values of the parameters,
the form of the Hamiltonian and the ratio between $\omega_{1,2}$ and $u_2$ fits the
description of superconducting transmon qubits, with qubit energies in the GHz
range and static qubit-qubit-coupling in the MHz range.

Introducing symmetries in the Hamiltonian~\eqref{eq:fullH} reduces the
controllability. First, we consider a situation in which
the two qubits operate at the same frequency $\omega_1 = \omega_2$. In this
case, the dynamic Lie algebra consists of only 9 instead of 15
operators. Consequently, not every
two-qubit gate can be implemented. However, the nine operators include
$\sigma_x^{(1)} \sigma_{x}^{(2)}$,
$\sigma_y^{(1)} \sigma_{y}^{(2)}$,
$\sigma_z^{(1)} \sigma_{z}^{(2)}$,
which are sufficient to reach every point in the Weyl chamber, cf.\
Eq.~\eqref{eq:cartan_decomp}. This is illustrated
on the right of Fig.~\ref{fig:fullControl}. Despite
the reduced controllability, the Weyl chamber is more evenly filled after the
same 1000 propagation steps as on the left. This counterintuitive
finding is due to the lower dimension of the random walk, with no
resources being ``wasted'' on the missing six single-qubit directions.

\begin{figure*}[tb]
  \centering
  \includegraphics{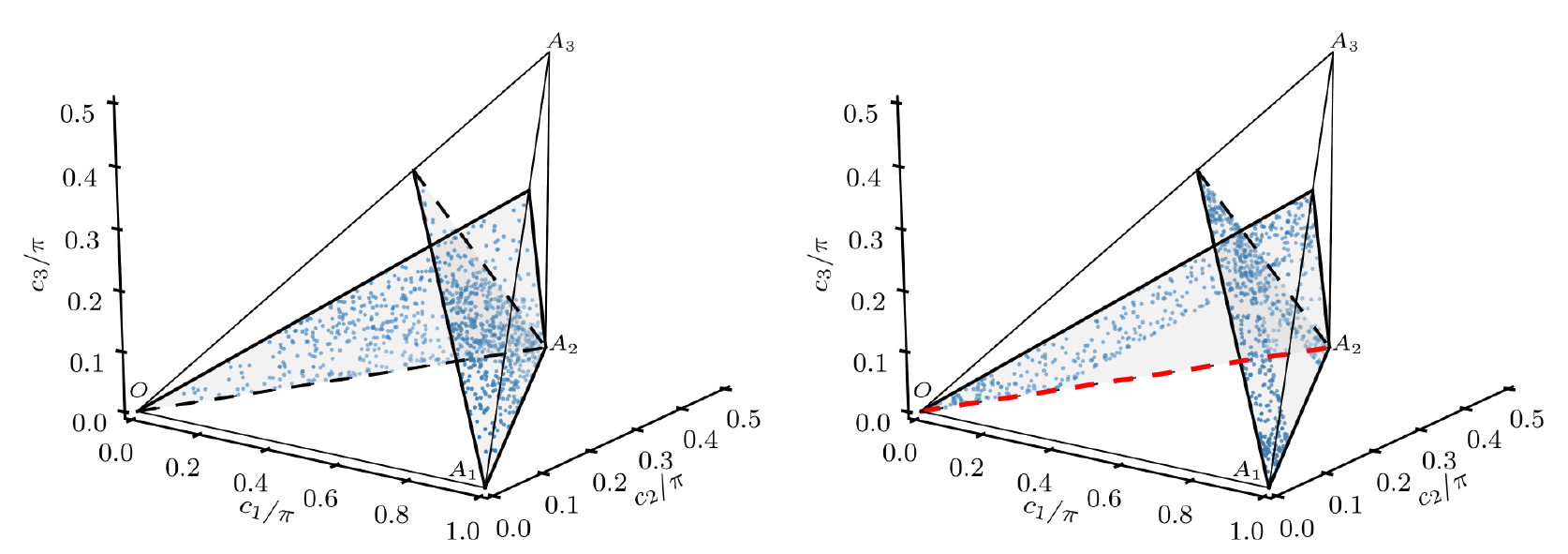}
  \caption{(Color online)
    Same as Fig.~\ref{fig:fullControl} for the fully degenerate case $\omega_1
    = \omega_2 = 0$ and two random pulses $u_1(t), u_2(t) \in [0,1]$.
  Not every point in the Weyl chamber can be reached. For independent
  pulses $u_1(t)$,   $u_2(t)$ (left),
  the dynamic Lie algebra consists of four generators, and a two-dimensional subset
  of the Weyl chamber can be reached, indicated by the shaded triangles,
  $O$--$(\frac{2\pi}{3},\frac{\pi}{3}, \frac{\pi}{3})$--$A_2$ and
  $A_1$--$(\frac{\pi}{3},\frac{\pi}{3}, \frac{\pi}{3})$--$A_2$. The reachable set
  is further reduced to a subset if
  $u_1(t) \equiv u_2(t)$ (right), i.e., the single-qubit and
  interaction operators couple to the same pulse.
  Lastly, without single-qubit driving ($u_1(t) \equiv 0$),
  only a one-dimensional subset of the Weyl chamber can be reached,
  the red line  $O$--$A_2$.}
  \label{fig:reducedControl}
\end{figure*}
The set of gates that can be implemented with Hamiltonian~\eqref{eq:fullH} is
more severely restricted if both qubits are
completely degenerate, $\omega_1 = \omega_2 = 0$. This is typical for
superconducting charge qubits operated at the  ``charge degeneracy
point''. Without any drift term, the Lie algebra consists of
only four generators,
$\sigma_z^{(1)} \sigma_{y}^{(2)} + \sigma_y^{(1)} \sigma_{z}^{(2)}$ and
$\sigma_y^{(1)} \sigma_{y}^{(2)} - \sigma_z^{(1)} \sigma_{z}^{(2)}$ in addition
to the two original terms.
The implications for controllability in the Weyl
chamber are not immediately obvious since three
generators can be sufficient to obtain full Weyl chamber
controllability. The
easiest approach is to perform a numerical analysis, the results of which are
shown on the left of Fig.~\ref{fig:reducedControl}. Two independent randomized
pulses $u_1(t)$ and $u_2(t)$ were used. The reachable points
lie on a plane, which due to the reflection symmetries
appears as two triangular branches. Note that almost none of the
common two-qubit gates are included in this set.

If only a single pulse is available to drive both the single-qubit and
two-qubit terms, $u_1(t) \equiv u_2(t)$, and the qubits are
degenerate, $\omega_1 = \omega_2 = 0$, there is a single
generator for the dynamics. This situation is shown
on the right of Fig.~\ref{fig:reducedControl}.
Although there is only a single generator for the dynamics, a
two-dimensional subset of the Weyl chamber can be reached. However,
the subset is no longer the full plane as it is for two independent
pulses (left of Fig.~\ref{fig:reducedControl}). Without single-qubit
control, the center of the plane is not longer reachable.
It is important to remember that while a single
generator yields points on a line in the Weyl chamber (not necessarily
a straight one), it can still fill
an arbitrary subset of the Weyl chamber, due to 
reflections at the boundaries.
A similar example, restricted to the ground plane of the
Weyl chamber, has been analyzed in Ref.~\cite{ZhangPRA03}.

Lastly, if there is no control over the individual qubits at all,
$u_1(t) \equiv 0$, the only remaining generator is
$\sigma_{x}^{(1)} \sigma_{x}^{(2)}+ \sigma_{y}^{(1)} \sigma_{y}^{(2)}$.
This corresponds to the straight line
$O$--$A_2$ in the Weyl chamber, shown in red in
Fig.~\ref{fig:reducedControl}. The line is reflected back onto itself
at the $A_2$ point. Thus, in this case only a truly one-dimensional subset
of reachable gates in the Weyl chamber can be realized.

For a Hamiltonian that allows for  a one-dimensional search-space only,
optimal control calculations with a functional targeting all perfect
entanglers will not yield results better than direct gate
optimization.
In contrast, for Hamiltonians allowing for two or three search directions in
the Weyl chamber, cf.\ Figures~\ref{fig:fullControl}
and~\ref{fig:reducedControl}, the polyhedron of perfect entanglers
may be approached from several different angles.  Optimization with a
functional targeting all perfect entanglers is then non-trivial. In
such a search, the optimized solution will depend on additional constraints
in the functional and the initial guess field. This will be explored
in the sequel to this paper.

\section{Summary}
\label{sec:concl}

We have revisited the parametrization of two-qubit gates, i.e.,
elements of the Lie group $SU(4)$, in terms of three real numbers, the
local invariants~\cite{ZhangPRA03}, in order to derive an optimization
functional for optimal control to target the whole subset of perfectly
entangling two-qubit gates.  We first identified an analytical
function of the local invariants $d(g_1,g_2,g_3)$ which becomes zero
at the boundary of the subset of perfect entanglers but can be of any
sign within this subset.  We rectified this ambiguity by using
$d(g_1,g_2,g_3)$ to obtain a functional $\mathcal{D}(U)$ that
determines definitively if we are within the set of perfect
entanglers.  Specifically, $\mathcal{D}(U)$ yields zero if a two-qubit gate $U$ is a
perfect entangler and is positive otherwise.

This functional
represents a generalization of our earlier work on optimizing for a
local equivalence class~\cite{MuellerPRA11} instead of a specific
gate~\cite{PalaoPRA03}.  Optimization with such a functional is useful
if one wants to implement an arbitrary perfect entangler.  In this
case, a functional targeting the whole subset of perfect entanglers
allows for more flexibility and thus potentially better control than
optimization for a specific gate or a single local equivalence class.
Furthermore, since gates locally equivalent to perfect entanglers
occupy nearly 85\% of $SU(4)$~\cite{WattsEnt13,MuszPRA13}, the target
of such a functional is very large indeed.

The full potential of such a generalized search strategy can, however,
only be utilized if the Hamiltonian is sufficiently complex, allowing
to approach the subset of perfect entanglers from more than one
direction.  For a generic two-qubit Hamiltonian, we have therefore
analyzed the basic requirements for a nontrivial search. Not
surprisingly, symmetries in the Hamiltonian preclude a full Weyl
chamber search. Caution is necessary in particular when operating in
the regime of the rotating-wave approximation which typically
introduces degeneracies and compromises complete controllability.

The sequel to this paper illustrates optimization with the perfect
entanglers' functional for several numerical examples.  The physical
models, when restricted to the logical subspace, correspond to the
generic Hamiltonian analyzed here.

\begin{acknowledgments}
We thank the Kavli Institute for
Theoretical Physics for hospitality and for supporting this 
research in part by the National Science Foundation Grant
No. PHY11-25915.
Financial support from the National Science Foundation under the
Catalzying International Collaborations program (Grant
No. OISE-1158954), the DAAD under grant PPP USA 54367416, the EC
through the  EU-IP projects SIQS and DIADEMS, the
DFG under SFB/TRR21 and the Science Foundation Ireland under Principal
Investigator Award 10/IN.1/I3013 is gratefully acknowledged.
\end{acknowledgments} 


\end{document}